# Optimal Allocation of Virtual Inertia and Droop Control for Renewable Energy in Stochastic Look-Ahead Power Dispatch

Yukang Shen, Wenchuan Wu, *Fellow, IEEE*, Shumin Sun, and Bin Wang

*Abstract*—To stabilize the frequency of the renewable energy sources (RESs) dominated power system, frequency supports are required by RESs through virtual inertia emulation or droop control in the newly published grid codes. Since the long-term RES prediction involves significant errors, we need online configure the frequency control parameters of RESs in a rolling manner to improve the operation economics under the premise of stabilizing system frequency. To address this concern, this paper proposes a frequency constrained stochastic look-ahead power dispatch (FCS-LAPD) model to formulate the frequency control parameters of RESs and Energy Storage Systems (ESSs) as scheduling variables, which can optimally allocate the virtual inertia and droop coefficient of RESs and ESSs. In this FCS-LAPD model, the uncertainties of RESs are characterized using Gaussian Mixture Model (GMM). The required reserves are determined by frequency control parameters, and the reserve cost coefficients are adjusted properly to allocate the reserves according to the predicted power generation. Due to the nonlinearity of the frequency nadir constraint, a convex hull approximation method is proposed to linearize it with guaranteed feasibility. The proposed FCS-LAPD is ultimately cast as an instance of quadratic programming and can be efficiently solved. Case studies on modified IEEE 24-bus system and a provincial power system in China are conducted to show the effectiveness of the proposed model.

*Index Terms*—Frequency constraint, look-ahead power dispatch, renewable energy, energy storage, parameter configuration, convex hull.

## NOMENCLATURE

### A. Abbreviations

| | |
|---|---|
| RES | Renewable energy source |
| ESS | Energy storage system |
| SoC | State of charge |
| PTDF | Power transfer distribution factor |
| RoCoF | Rate of change of frequency |

### B. Indices

| | |
|---|---|
| $t$ | Index of the time from 1 to $Nt$ |
| $i$ | Index of the thermal generator from 1 to $Ng$ |
| $j$ | Index of the RES from 1 to $Nw$ |
| $k$ | Index of the ESS from 1 to $Ne$ |
| $d$ | Index of the load from 1 to $Nd$ |
| $l$ | Index of the transmission line from 1 to $Nl$ |

### C. Parameters

| | |
|---|---|
| $a_i, b_i, c_i$ | Coefficients of the fuel cost function of generator $i$ |
| $rgc_i$ | Cost coefficient of reserve of generator $i$ |
| $P_i^{max}, P_i^{min}$ | Upper/lower bound of power output of generator $i$ |
| $Rup_i, Rdn_i$ | Ramp up/down rate limit of generator $i$ |
| $\beta_i$ | Affine regulation participation factor of generator $i$ |
| $H_i$ | Inertia constant of generator $i$ |
| $R_i$ | Droop coefficient of generator $i$ |
| $F_i$ | Fraction parameter of turbine of generator $i$ |
| $T_i$ | Governor-turbine time constant of generator $i$ |
| $rwc_j^t$ | Cost coefficient of reserve of RES $j$ at time $t$ |
| $rwc$ | Constant cost coefficient of RES |
| $W_{j,fore}^t$ | Forecast generation of RES $j$ at time $t$ |
| $\widehat{W}_j^t$ | Actual generation of RES $j$ at time $t$ (random variable) |
| $P_j^{cap}$ | Capacity of RES $j$ |
| $H_j^{max}$ | Upper bound of virtual inertia of RES $j$ |
| $D_j^{max}$ | Upper bound of droop coefficient of RES $j$ |
| $rec_k$ | Cost coefficient of reserve of ESS $k$ |
| $rec$ | Constant cost coefficient of ESS |
| $\eta_k^c, \eta_k^D$ | Charging/discharging efficiency of ESS $k$ |
| $P_k^{max}$ | Upper bound of charging/discharging power of ESS $k$ |
| $E_k^{max}, E_k^{min}$ | Upper/lower bound of SoC of ESS $k$ |
| $H_k^{max}$ | Upper bound of virtual inertia of ESS $k$ |
| $D_k^{max}$ | Upper bound of droop coefficient of ESS $k$ |
| $D_d^t$ | Power demand of load $d$ at time $t$ |
| $D_0$ | Load damping coefficient |
| $\Delta P^t$ | Imaginary power disturbance at time $t$ |
| $P_{max}^l$ | Power capacity of transmission line $l$ |
| $s_i^l, s_j^l, s_k^l, s_d^l$ | PTDF of transmission line $l$ |
| $s_{j,aff}^l$ | Equivalent PTDF of RES $j$ considering affine regulation |

*This work was supported in part by the National Key Research and Development Plan of China under Grant 2022YFBXXXX and in part by the National Science Foundation of China under Grant 51725703. (Corresponding author: Wenchuan Wu)

Y. Shen, W. Wu and B. Wang are with the State Key Laboratory of Power Systems, Department of Electrical Engineering, Tsinghua University, Beijing 100084, China (e-mail: syk21@mails.tsinghua.edu.cn; wuwench@tsinghua.edu.cn; wb1984@tsinghua.edu.cn).

S. Sun is with the Electric Power Research Institute of State Grid Shandong Electric Power Co., Ltd., Jinan, Shandong Province 250003, China (epri@163.com).

| $\alpha_{\mathrm{Gu}}, \alpha_{\mathrm{Gd}}$ | Maximum allowable probability of upward/downward power over-bound |
| $\alpha_{\mathrm{Rw}}$ | Maximum allowable probability of reserve insufficiency |
| $\alpha_{\mathrm{L+}}, \alpha_{\mathrm{L-}}$ | Maximum allowable probability of transmission line overloading (bidirectional) |
| $f_0$ | Nominal frequency |
| $f_{max}^{lim}$ | Maximum allowable frequency deviation |
| $\dot{f}^{lim}$ | Maximum allowable RoCoF |
| $f_{ss}^{lim}$ | Maximum allowable steady-state frequency deviation |
| $P_{base}$ | Base power of the power system |

*D. Variables*

| $FC_i^t$ | Fuel cost of generator $i$ at time $t$ |
| $RGC_i^t$ | Reserve cost of generator $i$ at time $t$ |
| $P_i^t$ | Power output of generator $i$ at time $t$ |
| $Rg_i^t$ | Reserve of generator $i$ at time $t$ |
| $RWC_j^t$ | Reserve cost of RES $j$ at time $t$ |
| $W_{j,sche}^t$ | Scheduled power of RES $j$ at time $t$ |
| $Rw_j^t$ | Reserve of RES $j$ at time $t$ |
| $H_j^t$ | Virtual inertia of RES $j$ at time $t$ |
| $D_j^t$ | Droop coefficient of RES $j$ at time $t$ |
| $Loss_k^t$ | Power loss of ESS $k$ at time $t$ |
| $REC_k^t$ | Reserve cost of ESS $k$ at time $t$ |
| $LossC_k^t$ | Charging power loss of ESS $k$ at time $t$ |
| $LossD_k^t$ | Discharging power loss of ESS $k$ at time $t$ |
| $P_k^t$ | Charging/discharging power of ESS $k$ at time $t$ |
| $Re_k^t$ | Reserve of ESS $k$ at time $t$ |
| $E_k^t$ | SoC of ESS $k$ at time $t$ |
| $H_k^t$ | Virtual inertia of ESS $k$ at time $t$ |
| $D_k^t$ | Droop coefficient of ESS $k$ at time $t$ |
| $H_{sys}^t$ | Total inertia of the power system at time $t$ |
| $D_{sys}^t$ | Total damping of the power system at time $t$ |

# I. INTRODUCTION

## A. Background

THE penetration of renewable energy brings challenges to maintain frequency stability of power systems. The inherent properties of RESs such as uncertainty, intermittency, fluctuation, and low inertia increase the probability of power disturbances and may cause frequency instability after a contingency. With more and more conventional generators replaced by RESs, the frequency supports would be insufficient only from conventional generators, so that the RES units such as wind turbines and photovoltaics are required to provide frequency regulation service in the newly published grid codes [1][2]. The RES units can emulate the inertia response and governor behavior of synchronous units [3], which are called virtual inertia control and droop control.

The frequency dynamics during power disturbances are affected by all resources participating in primary frequency regulation, so that the frequency regulation resources should be coordinated and allocated by system operators. Frequency constrained unit commitment (UC) has been proposed to schedule thermal generators to maintain frequency security [4]-[7]. The control parameters of RESs such as virtual inertia and droop coefficient were fixed in [4]-[7]. However, in RES dominated power systems, fixed control parameters could not make trade-off between the security and economics, since the required frequency supports vary according to the load level during one day. Therefore, these frequency control parameters should be online determined according to the realistic conditions in different scheduling periods [8].

For RES units, the frequency control parameters should match the regulation reserves, which incur power curtailment and are strongly coupled with the power dispatch, so that the RESs' frequency regulation capability should be optimized coordinated with the power scheduling. Since the long-term RES prediction involves significant errors in UC, which is usually solved in day-ahead or week-ahead stage, the RESs' frequency control parameters should be optimized online in short-term time stage, i.e. in look-ahead power dispatch.

## B. Literature Review

The premise of designing the frequency control parameters of RESs is to model the frequency dynamics of the power systems. There are mainly two solutions of frequency dynamics modeling. The first approach uses the swing equation with the assumption that mechanical power is piecewise linear respect to time [9]-[10]. This approach decouples the mechanical power and frequency, which does not comply with realistic frequency control system. The second approach derives the frequency dynamics based on the low-order system frequency response model [11]. For multi-machine systems with high-order frequency model, it is difficult to obtain the analytical formulation of frequency trajectory [12]. A low-order model is obtained by aggregating multi machines into single machine [13]-[14]. Due to the fast electromagnetic transients in power electronic interfaces, time delay of the frequency regulation from RESs could be ignored [15], then the frequency support from RESs can be included in frequency dynamics model and an analytical formulation of frequency trajectory can be derived.

Many studies have focused on the optimal allocation of virtual inertia and droop control for RESs. The inertia and damping distribution were obtained by minimizing the system norm in [16]-[18]. The Lyapunov function in system norm optimization make it non-convex, and the gradient algorithm is employed to solve it. In [19], reinforcement learning method was used to obtain the optimal control strategy for virtual synchronous generators.

Some other approaches have been explored based on the frequency stability point of view. The frequency dynamic metrics such as frequency nadir, steady-state frequency, and RoCoF should be kept in allowable range, and the difficulty is how to deal with the nonlinear frequency nadir constraint. The iterative, sensitivity-based optimization algorithms were proposed in [20]-[21]. The sensitivity with respect to inertia and damping were analyzed, and the optimal allocation of inertia and damping were obtained based on an iterative process. In [22], the optimal participation of distributed energy resources in inertial- and primary-frequency response were studied. Each RES shares the inertia and damping in proportion to its power rating. In [23], a chance-constrained unit commitment was proposed to provide the optimal allocation of virtual inertia in wind-storage systems. The total synthetic inertia was online optimized in microgrid scheduling in [24] while without considering the droop response.

Energy storage systems have been used to provide frequency regulation service [25]. The optimal placement of virtual inertia for ESSs was proposed in [26] based on frequency stability point of view. To maintain system reliability, the newly published grid codes have specific requirements on RESs to provide ancillary services [1][2] for power systems with high penetration of RESs. Since the participation of ancillary service incurs power loss, the RESs' frequency regulation capability should be optimized online accompanied with the look-ahead power dispatch. This concern has not been addressed in previous literatures.

### C. Contributions

In this paper, a frequency constrained stochastic look-ahead power dispatch (FCS-LAPD) model is proposed where the virtual inertia and droop coefficient of RESs/ESSs are regarded as scheduling variables to be optimized along with online power scheduling. The main contributions are summarized as follows:

1) The allocation of virtual inertia and droop coefficient determines the required reserves of RESs/ESSs, which are strongly coupled with the power dispatch. Therefore, we propose a FCS-LAPD model to comprehensively optimize the scheduling power and reserve for all units, and allocate the virtual inertia and droop coefficient of RESs/ESSs. The uncertainties of RESs are characterized using Gaussian Mixture Model (GMM), and the reserve is allocated to each RES/ESS in proportion to its forecast generation/power capability.

2) To make the proposed FCS-LAPD model tractable, the nonlinear frequency nadir constraint is transformed into a half-space intersection formulation by employing the proposed convex hull approximation method. Since the convexity of the frequency nadir constraint is proved, this reformulation can guarantee the feasibility of the original frequency nadir constraint and its conservativeness can be adjusted. The proposed FCS-LAPD model is ultimately cast as an instance of quadratic programming and can be efficiently solved.

The rest of this paper is organized as follows: Section II formulates the frequency dynamics and derives the frequency security constraints. Section III develops the FCS-LAPD model. Case studies and results are demonstrated in Section IV. Conclusions are drawn in Section V.

## II. FREQUENCY DYNAMICS MODELING AND FREQUENCY CONSTRAINTS DERIVATION

This section describes the frequency dynamics modeling and three metrics: frequency nadir, steady-state frequency, and RoCoF. Then, the frequency constraints and their convex transformation are derived.

### A. Frequency Dynamics Modeling

The frequency dynamics under a disturbance are mostly influenced by the primary frequency regulation of all online units. By aggregating all individual generators into a single machine [13]-[15], the frequency dynamics could be modeled using one swing equation. The aggregated frequency is also called Center of Inertia (COI) frequency. Considering a power system comprised of thermal generators, RES units, and ESSs, the frequency response could be modeled as in Fig. 1.

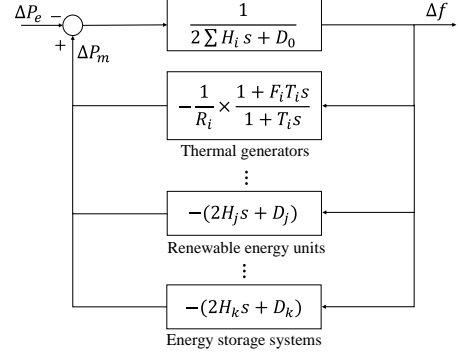

Fig. 1. System frequency response model.

For thermal generators with reheat steam turbines, the governor-turbine transfer function can be formulated as in Fig.1, and for those with non-reheat steam turbines, we can set $F_i = 0$ for a uniform formulation.

The RES units can provide frequency support by operating in de-loading mode or with installed ESSs. In this paper, we separate the RES units and ESSs from the hybrid RES-ESS station to two individual units, then the RES units can only provide frequency support by degenerating their outputs, and the ESSs can be independent to participate in power dispatch for peak shaving. Compared with conventional thermal generators, the electromagnetic transients in power electronic devices are fast enough [15], so that the time delay of the frequency regulation from RESs and ESSs could be omitted. The control strategies for RESs and ESSs contain virtual inertia control and droop control, as shown in Fig. 1.

The high-order frequency response model in Fig. 1 can be transformed into a low-order model under the assumption that $T_i = T$ [13]-[14]. The transfer function can be derived as:

$$\frac{\Delta f(s)}{\Delta P_e(s)} = -\frac{1}{2HT} \times \frac{1 + Ts}{s^2 + 2\zeta\omega_n s + \omega_n^2} \quad (1a)$$

where

$$H = \sum_{i=1}^{Ng} H_i \frac{P_i^{max}}{P_{base}} + \sum_{j=1}^{Nw} H_j \frac{P_j^{cap}}{P_{base}} + \sum_{k=1}^{Ne} H_k \frac{P_k^{max}}{P_{base}} \quad (1b)$$

$$D = D_0 + \sum_{j=1}^{Nw} D_j \frac{P_j^{cap}}{P_{base}} + \sum_{k=1}^{Ne} D_k \frac{P_k^{max}}{P_{base}} \quad (1c)$$

$$F = \sum_{i=1}^{Ng} \frac{F_i}{R_i} \frac{P_i^{max}}{P_{base}}, \qquad R = \sum_{i=1}^{Ng} \frac{1}{R_i} \frac{P_i^{max}}{P_{base}} \quad (1d)$$

$$\omega_n = \sqrt{\frac{D+R}{2HT}}, \qquad \zeta = \frac{2H + (D+R)T}{2\sqrt{2HT(D+R)}} \quad (1e)$$

The natural frequency $\omega_n$ and damping ratio $\zeta$ are functions on aggregated inertia $H$, aggregated damping $D$, and thermal generator aggregated parameters $R$, $F$.

We assume the power disturbances are stepwise $\Delta P_e(s) = \Delta P/s$, then the time domain expression of frequency deviation can be derived as well as three metrics: RoCoF, steady-state frequency, and frequency nadir:

$$\dot{f}_{max} = \frac{f_0 \Delta P}{2H} \tag{2a}$$

$$\Delta f_{ss} = \frac{f_0 \Delta P}{R + D} \tag{2b}$$

$$\Delta f_{max} = \frac{f_0 \Delta P}{R + D}\left(1 + e^{-\zeta \omega_n t_{nadir}}\sqrt{\frac{T(R - F)}{2H}}\right) \tag{2c}$$

where $\dot{f}_{max}$ represents the maximum RoCoF, $\Delta f_{ss}$ is the steady-state frequency deviation, and $\Delta f_{max}$ is the maximum frequency deviation. $t_{nadir} = 1/\omega_r \tan^{-1}[\omega_r/(\zeta\omega_n - T^{-1})]$ is the time when frequency reaches nadir with $\omega_r = \omega_n\sqrt{1 - \zeta^2}$.

The power disturbance $\Delta P$ should be adjusted by system operators to make the trade-off between security and economy. This disturbance can be set as the maximum output of all units to simulate the $N - 1$ case or varies according to the load level/RES generation to simulate the fluctuation of load/RESs.

### B. Frequency Constraints Derivation

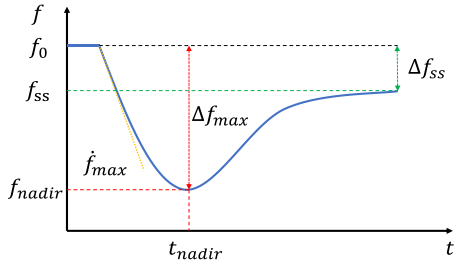

Fig. 2. Frequency dynamics after power disturbance.

Fig. 2 shows the frequency dynamics after a disturbance. The three metrics: RoCoF, steady-state frequency, and frequency nadir should be kept in allowable range, as the constraints in scheduling period:

$$\frac{f_0 \Delta P}{2H} \leq \dot{f}^{lim} \tag{3a}$$

$$\frac{f_0 \Delta P}{R + D} \leq f_{ss}^{lim} \tag{3b}$$

$$\frac{f_0 \Delta P}{R + D}\left(1 + e^{-\zeta \omega_n t_{nadir}}\sqrt{\frac{T(R - F)}{2H}}\right) \leq f_{max}^{lim} \tag{3c}$$

To provide primary frequency regulation, generators, RESs and ESSs should also keep adequate reserves, as follows:

$$Rg_i \geq \frac{1}{R_i}P_i^{max}\frac{f_{ss}^{lim}}{f_0} \tag{4a}$$

$$Rw_j \geq D_j P_j^{cap}\frac{f_{max}^{lim}}{f_0} + 2H_j P_j^{cap}\frac{\dot{f}^{lim}}{f_0} \tag{4b}$$

$$Re_k \geq D_k P_k^{max}\frac{f_{max}^{lim}}{f_0} + 2H_k P_k^{max}\frac{\dot{f}^{lim}}{f_0} \tag{4c}$$

Constraint (4a) requests generators to keep adequate reserve when frequency reaches steady-state. For RESs and ESSs, the reserve contains two parts: one part provides virtual inertia and the other provides droop control [27].

To obtain the optimal allocation of virtual inertia and droop control for RES, the total inertia $H$ and total damping $D$ should be determined under the frequency constraints (3) in each schedule period. The RoCoF constraint (3a) and the steady-state frequency constraint (3b) are convex on $H$ and $D$. However, the frequency nadir constraint (3c) is nonlinear, which will make the optimization model intractable. In this paper, we propose the convex hull approximation (CHA) method to transform the nonlinear constraint (3c) into a convex linear formulation.

Firstly, the convexity of the maximum frequency deviation $\Delta f_{max}$ on $H$ and $D$ in actual power systems is proved. (Proof is provided in Appendix I)

Then, the feasible region of constraint (3c) corresponds to the $f_{max}^{lim}$-sublevel set of $\Delta f_{max}$. According to the properties of convex function, the feasible region of constraint (3c) is a convex set. In 2-dimensional space, a bounded closed convex set can be approximated by a convex polygon, so we propose the convex hull approximation method to approximate the feasible region of (3c), and the solution process is as follows:

1) Perform Monte Carlo sampling within the upper and lower bounds of total inertia $H$ and total damping $D$ to obtain initial data samples.
2) Calculate the maximum frequency deviation $\Delta f_{max}$ corresponding to each sample based on (2c), and select the train samples that satisfy constraint (3c).

$$\mathbf{X_s} = \begin{bmatrix} H_1 & H_2 & H_3 & \cdots & H_s & \cdots \\ D_1 & D_2 & D_3 & \cdots & D_s & \cdots \end{bmatrix} \tag{5}$$

   where $\mathbf{X_s}$ represent the train samples that satisfy the constraint (3c), and $H_s$ and $D_s$ represent the inertia and damping of sample $s$, respectively.
3) Solve the convex hull of the samples $\mathbf{X_s}$ using Quickhull algorithm and get its half-space intersection formulation.

$$\omega_p^H H + \omega_p^D D + b_p \geq 0, p = 1, 2, \cdots, P \tag{6}$$

   where $\omega_p^H$, $\omega_p^D$ represent the normal vector element of the $p$-th hyperplane, and $b_p$ is the offset. $P$ is the total number of hyperplanes.

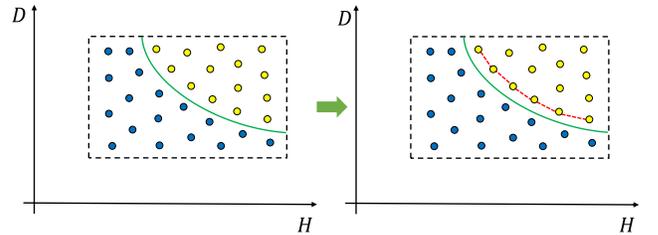

Fig. 3. Convex hull approximation.

Fig. 3 shows the convex hull approximation for constraint (3c). The lower bound of total inertia $H$ in step 1) represents the case when RES and ESS do not provide virtual inertia, and the upper bound of $H$ means the virtual inertia from each RES/ESS reach its upper bound. The same is with total damping $D$. The yellow point in Fig. 3 represents the sample satisfies (3c), while the blue one represents the sample that does not satisfy. The red dashed lines in Fig. 3 represent the hyperplanes in (6) and the green solid line is the boundary of the nonlinear constraint (3c).

The train sample $\mathbf{X_s}$ in step 2) is a finite point set and its convex hull is a bounded convex polygon. To solve the convex hull of $\mathbf{X_s}$, we use the Quickhull algorithm [29]. It can return the extreme points for a finite point set by employing two main operations: oriented hyperplane through $d$ points and signed distance to hyperplane.

To embed it into optimization model, the convex polygon should be represented by half-space intersection (6), and each hyperplane can be determined by two adjacent vertices. Based on the proposed CHA method, the nonlinear constraint (3c) can be transformed into a group of linear constraints (6).

*Remark 1*: The convex hull approximation method is the convex relaxation for the train samples $X_s$, not for the original constraint (3c). The CHA process is equivalent to searching the extreme points of $X_s$ on the boundary (green solid line in Fig. 3) or the closest to the boundary, and then approximating the boundary with multi secant lines. The approximated constraint (6) is conservative because (6) is the smallest convex set that contains $X_s$, so (6) is the subset of (3c). The conservativeness can be reduced by increasing the number of samplings.

TABLE I
THE IMPACT OF DIFFERENT SAMPLE SIZES ON CHA METHOD

| The size of initial samples | 10000 | 20000 | 50000 |
|---|---|---|---|
| The number of hyperplanes | 6 | 7 | 8 |
| Computational time (s) | 0.007 | 0.011 | 0.016 |
| Classification error (%) | 0.07 | 0.04 | 0.02 |

The CHA method is solved with different initial sample sizes, and the results are shown in Table I. The obtained approximated constraints (6) can be regarded as a binary classification model. The classification error is the ratio of the number of misclassified samples to the total number of test samples ($10^4$). It is shown that the classification error decreases with the size of samples increasing. Due to the conservativeness of the CHA method, only a few "safe samples" (yellow points in Fig. 3) might be misclassified, while no "unsafe samples" (blue points in Fig. 3) are misclassified. This conservativeness can guarantee the feasibility of the original frequency nadir constraint (3c), which is the advantage compared to other methods such as piecewise linearization (PWL), as shown in Fig. 4. In addition, the computation burden of the CHA method is small and can be solved online.

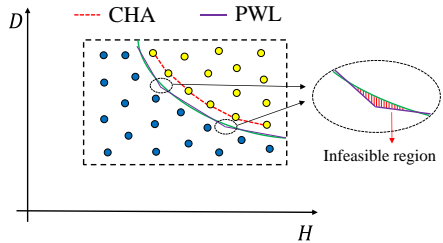

Fig. 4. Comparison between the CHA method and PWL method.

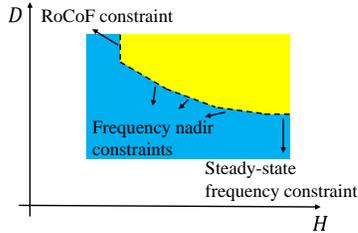

Fig. 5. Feasible region of frequency constraints (3).

Fig. 5 shows the feasible region (yellow region) of frequency constraints (3). The approximated frequency nadir constraints are a group of linear constraints (6), and the RoCoF constraint (3a) and steady-state frequency constraint (3b) are axis-parallel.

## III. FREQUENCY CONSTRAINED STOCHASTIC LOOK-AHEAD POWER DISPATCH

This section develops a frequency constrained stochastic look-ahead power dispatch model to provide the optimal allocation of virtual inertia and droop control for RESs/ESSs.

### A. Modelling of RES Uncertainty

Gaussian Mixture Model is used to model uncertainties of RESs [30]-[31], the probability density function is defined as:

$$\text{PDF}_X(x) = \sum_{m=1}^{M} \alpha_m N_m(x, \boldsymbol{\mu_m}, \boldsymbol{\Sigma_m}) \tag{7a}$$

$$\sum_{m=1}^{M} \alpha_m = 1, \qquad \alpha_m > 0 \tag{7b}$$

where $N_m(\cdot)$ denotes the $m$-th multi-dimensional Gaussian distribution, and $\alpha_m$, $\boldsymbol{\mu_m}$, $\boldsymbol{\Sigma_m}$ represent the weight coefficient, expectation vector and covariance matrix, respectively.

The joint probabilistic distribution of actual RES generation can be approximated by GMM and the parameters $\alpha_m$, $\boldsymbol{\mu_m}$, $\boldsymbol{\Sigma_m}$ can be determined by fitting historical data [31].

### B. Objective Function

The objective of the FCS-LAPD optimization is to minimize the operating cost, which consists of fuel cost, reserve cost, and ESS power losses.

$$\min \sum_{t=1}^{Nt} [\sum_{i=1}^{Ng} (FC_i^t + RGC_i^t) + \sum_{j=1}^{Nw} RWC_j^t + \sum_{k=1}^{Ne} (Loss_k^t + REC_k^t)] \tag{8}$$

*1) Fuel Cost:*

$$FC_i^t = a_i(P_i^t)^2 + b_i P_i^t + c_i \tag{9}$$

The fuel cost of thermal generators can be represented by quadratic function of the power output $P_i^t$.

*2) Reserve Cost of Generators:*

$$RGC_i^t = rgc_i Rg_i^t \tag{10}$$

The reserve cost of thermal generators is proportional to the reserve capacity $Rg_i^t$.

*3) Reserve Cost of RES:*

$$RWC_j^t = rwc_j^t (Rw_j^t)^2 \tag{11a}$$

$$rwc_j^t = \frac{rwc}{W_{j,fore}^t} \tag{11b}$$

The reserve cost of RES is proportional to square of the reserve $Rw_j^t$, which is similar to the curtailment penalty in [31]. The cost coefficient $rwc_j^t$ is inversely proportional to the forecast generation $W_{j,fore}^t$, which can ensure that each RES unit shares the reserve in proportion to its forecast generation with minimal curtailment. (Proof is provided in Appendix II)

*4) ESS Power Losses:*

$$Loss_k^t = \max(LossD_k^t, LossC_k^t) \tag{12}$$

$$LossD_k^t = \left(\frac{1}{\eta_k^D} - 1\right) P_k^t \tag{13a}$$

$$LossC_k^t = (\eta_k^C - 1) P_k^t \tag{13b}$$

The ESS power losses come from the charging/discharging process. $P_k^t > 0$ denotes the discharging power, in this case $LossD_k^t > 0$, $LossC_k^t < 0$, so $Loss_k^t = LossD_k^t$; and $P_k^t < 0$ represents charging power, in this case $LossD_k^t < 0$, $LossC_k^t >$ 0, so $Loss_k^t = LossC_k^t$. The constraint (12) is non-convex, and its relaxation will be discussed in the next subsection.

*5) Reserve Cost of ESS:*

$$REC_k^t = rec_k(Re_k^t)^2 \tag{14a}$$

$$rec_k = \frac{rec}{P_k^{max}} \tag{14b}$$

The cost function of ESS reserve is similar to that of RES. The cost coefficient $rec_k$ is inversely proportional to the power rating $P_k^{max}$, so that each ESS will share the reserve in proportion to its power rating.

## C. Constraints

*1) Power Balance Constraints:*

$$\sum_{i=1}^{Ng} P_i^t + \sum_{j=1}^{Nw} W_{j,sche}^t + \sum_{k=1}^{Ne} P_k^t = \sum_{d=1}^{Nd} D_d^t \tag{15}$$

The sum of generation from all units equals to the sum of power load to maintain the system power balance.

*2) Generator Constraints:*

$$P_i^t + Rg_i^t \le P_i^{max} \tag{16a}$$

$$P_i^t \ge P_i^{min} \tag{16b}$$

$$P_i^t - P_i^{t-1} \le Rup_i \tag{17a}$$

$$P_i^{t-1} - P_i^t \le Rdn_i \tag{17b}$$

For thermal generators, constraints (16) provide the upper and lower bounds on power output. Upward and downward ramping rates are constrained in (17).

*3) RES Constraints:*

$$W_{j,fore}^t = E(\widetilde{W}_j^t) \tag{18}$$

$$W_{j,sche}^t + Rw_j^t = W_{j,fore}^t \tag{19a}$$

$$0 \le W_{j,sche}^t \le W_{j,fore}^t \tag{19b}$$

The forecast generation $W_{j,fore}^t$ equals to the expectation of the actual generation $\widetilde{W}_j^t$, where $\widetilde{W}_j^t$ is a random variable and its distribution can be represented by a GMM. The relationship among the scheduled power $W_{j,sche}^t$, reserve $Rw_j^t$, and forecast generation $W_{j,fore}^t$ are shown in (19).

*4) ESS Constraints:*

$$E_k^t = E_k^{t-1} - (P_k^t + Loss_k^t)\Delta t \tag{20a}$$

$$E_k^{min} \le E_k^t \le E_k^{max} \tag{20b}$$

$$-P_k^{max} \le P_k^t \le P_k^{max} \tag{20c}$$

$$P_k^t + Re_k^t \le P_k^{max} \tag{20d}$$

$$Loss_k^t \ge LossD_k^t \tag{20e}$$

$$Loss_k^t \ge LossC_k^t \tag{20f}$$

The SoC constraints are shown in (20a)-(20b), where $\Delta t$ is the scheduling time period. The charging/discharging power is constrained in (20c)-(20d). (20e)-(20f) are the relaxation of (12) [32]. Due to the existence of $Loss_k^t$ in objective function, the optimal $Loss_k^t$ will equal to the larger one of $LossD_k^t$ and $LossC_k^t$.

The required energy of ESS to provide frequency support comes from its own stored energy. To guarantee a sufficient energy storage, we can set $E_k^{min} = P_k^{max} \times \Delta t_{PFR}$, where $\Delta t_{PFR}$ is the time for ESS to support primary frequency regulation.

*5) Generator Constraints Considering Affine Regulation:*

$$\bar{P}_i^t = P_i^t - \beta_i \sum_{j=1}^{Nw} (\widetilde{W}_j^t - Rw_j^t - W_{j,sche}^t) \tag{21a}$$

$$\Pr(\bar{P}_i^t + Rg_i^t \le P_i^{max}) \ge 1 - \alpha_{Gu} \tag{21b}$$

$$\Pr(\bar{P}_i^t \ge P_i^{min}) \ge 1 - \alpha_{Gd} \tag{21c}$$

For generators with nonzero participation factor $\beta_i$, the actual output $\bar{P}_i^t$ is regulated affinely in response to the forecast uncertainty of RESs, as shown in (21a). The upward/downward power limit constraints (21b)-(21c) are chance constraints [33], where $\Pr(\cdot)$ is the probability.

*6) Frequency Constraints:*

$$H_{sys}^t = \frac{1}{P_{base}} \left( \sum_{i=1}^{Ng} H_i P_i^{max} + \sum_{j=1}^{Nw} H_j^t P_j^{cap} + \sum_{k=1}^{Ne} H_k^t P_k^{max} \right) \tag{22a}$$

$$D_{sys}^t = D_0 + \frac{1}{P_{base}} \left( \sum_{j=1}^{Nw} D_j^t P_j^{cap} + \sum_{k=1}^{Ne} D_k^t P_k^{max} \right) \tag{22b}$$

$$H_{sys}^t \ge \frac{f_0 \Delta P^t}{2 \dot{f}^{lim}} \tag{22c}$$

$$D_{sys}^t \ge \frac{f_0 \Delta P^t}{f_{ss}^{lim}} - \frac{1}{P_{base}} \sum_{i=1}^{Ng} \frac{1}{R_i} P_i^{max} \tag{22d}$$

$$\omega_p^H H_{sys}^t + \omega_p^D D_{sys}^t + b_p \ge 0, p = 1,2,\cdots,P \tag{22e}$$

$$0 \le H_j^t \le H_j^{max}, \qquad 0 \le D_j^t \le D_j^{max} \tag{22f}$$

$$0 \le H_k^t \le H_k^{max}, \qquad 0 \le D_k^t \le D_k^{max} \tag{22g}$$

The total inertia and total damping are defined in (22a)-(22b). The RoCoF constraint (22c) and the steady-state frequency constraint (22d) are linear constraints on $H_{sys}^t$ and $D_{sys}^t$. The frequency nadir constraint (22e) is derived by employing CHA method proposed in Section II. The allowable tuning range for virtual inertia and droop coefficient of RESs/ESSs are constrained in (22f)-(22g). These upper bounds of frequency control parameters are determined by each RES/ESS considering their technical capacities and physics, and should be submitted to the system operators from each RES/ESS.

*7) Reserve Constraints:*

$$Rg_i^t \ge \frac{1}{R_i} P_i^{max} \frac{f_{ss}^{lim}}{f_0} \tag{23}$$

$$Re_k^t \ge D_k^t P_k^{max} \frac{f_{max}^{lim}}{f_0} + 2H_k^t P_k^{max} \frac{\dot{f}^{lim}}{f_0} \tag{24}$$

$$\Pr\left( \widetilde{W}_j^t - W_{j,sche}^t \ge D_j^t P_j^{cap} \frac{f_{max}^{lim}}{f_0} + 2H_j^t P_j^{cap} \frac{\dot{f}^{lim}}{f_0} \right) \ge 1 - \alpha_{Rw} \tag{25}$$

The reserve constraints for thermal generators and ESSs are the same with (4) in Section II. However, due to the forecasting uncertainty of RESs, the reserve constraint for RESs (25) should be chance constraint, which indicates that each RES unit can provide sufficient reserve under $1 - \alpha_{Rw}$ confidence level.

*8) Transmission Line Constraints:*

$$\Pr\left( \begin{array}{l} \sum_{i=1}^{Ng} s_i^l \bar{P}_i^t + \sum_{j=1}^{Nw} s_j^l (\widetilde{W}_j^t - Rw_j^t) \\ + \sum_{k=1}^{Ne} s_k^l P_k^t - \sum_{d=1}^{Nd} s_d^l D_d^t \le P_{max}^l \end{array} \right) \ge 1 - \alpha_{L+} \tag{26a}$$

$$\Pr\left( \begin{array}{l} \sum_{i=1}^{Ng} s_i^l \bar{P}_i^t + \sum_{j=1}^{Nw} s_j^l (\widetilde{W}_j^t - Rw_j^t) \\ + \sum_{k=1}^{Ne} s_k^l P_k^t - \sum_{d=1}^{Nd} s_d^l D_d^t \ge -P_{max}^l \end{array} \right) \ge 1 - \alpha_{L-} \tag{26b}$$

The power flow of transmission lines is stochastic due to the uncertainty of RESs, so the transmission line constraints (26)

are also chance constraints to ensure the bidirectional line flow security under predefined confidence level.

### D. Deterministic Transformation of Chance Constraints

According to the affine invariance of GMM [31][33], the chance constraints (21), (25), and (26) can be transformed into equivalent linear constraints with notation of quantile where $Q(\xi|\alpha)$ denotes the $\alpha$-quantile of random variable $\xi$.

$$\frac{P_i^t + Rg_i^t - P_i^{max}}{\beta_i} + \sum_{j=1}^{Nw}\left(Rw_j^t + W_{j,sche}^t\right) \le Q\left(\sum_{j=1}^{Nw}\widehat{W}_j^t \,|\alpha_{Gu}\right) \quad (27a)$$

$$\frac{P_i^t - P_i^{min}}{\beta_i} + \sum_{j=1}^{Nw}\left(Rw_j^t + W_{j,sche}^t\right) \ge Q\left(\sum_{j=1}^{Nw}\widehat{W}_j^t \,|1-\alpha_{Gd}\right) \quad (27b)$$

$$W_{j,sche}^t + D_j^t P_j^{cap}\frac{f_{max}^{lim}}{f_0} + 2H_j^t P_j^{cap}\frac{\dot{f}^{lim}}{f_0} \le Q(\widehat{W}_j^t|\alpha_{Rw}) \quad (28)$$

$$P_{max}^l - \sum_{i=1}^{Ng} s_i^l P_i^t + \sum_{j=1}^{Nw} s_{j,aff}^l Rw_j^t - M^l \sum_{j=1}^{Nw} W_{j,sche}^t$$
$$-\sum_{k=1}^{Ne} s_k^l P_k^t + \sum_{d=1}^{Nd} s_d^l D_d^t \ge Q\left(\sum_{j=1}^{Nw} s_{j,aff}^l \widehat{W}_j^t \,|1-\alpha_{L+}\right) \quad (29a)$$

$$P_{max}^l + \sum_{i=1}^{Ng} s_i^l P_i^t - \sum_{j=1}^{Nw} s_{j,aff}^l Rw_j^t + M^l \sum_{j=1}^{Nw} W_{j,sche}^t$$
$$+\sum_{k=1}^{Ne} s_k^l P_k^t - \sum_{d=1}^{Nd} s_d^l D_d^t \ge Q\left(\sum_{j=1}^{Nw} -s_{j,aff}^l \widehat{W}_j^t \,|1-\alpha_{L-}\right) \quad (29b)$$

$$s_{j,aff}^l = s_j^l - \sum_{i=1}^{Ng} s_i^l \beta_i, \qquad M^l = \sum_{i=1}^{Ng} s_i^l \beta_i \quad (29c)$$

In summary, the formulations (8)-(11), (13)-(20), (22)-(24), (27)-(29) compose the proposed FCS-LAPD model which is a quadratic programming (QP) problem. The Newton method can be employed to solve the quantile $Q(\xi|\alpha)$ [31][31]. The calculation process for quantile and the CHA process for frequency nadir constraint are all solved before solving the FCS-LAPD model.

## IV. CASE STUDIES

The framework of the proposed FCS-LAPD model is shown in Fig. 6. The system operators launch the FCS-LAPD model in a receding horizon every 1 hour to determine the power output and regulation reserve for all units in the upcoming 4 h, with a time resolution of 15 min. As for RESs/ESSs, the frequency control parameters (virtual inertia and droop coefficient) are also allocated to them. Each look-ahead power dispatch calculation consists of 16 points, and only the first 4 points would be put into control.

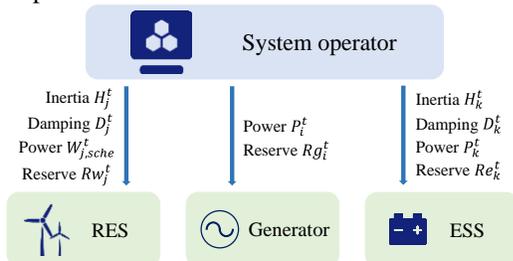

Inertia $H_j^t$
Damping $D_j^t$
Power $W_{j,sche}^t$
Reserve $Rw_j^t$

Power $P_i^t$
Reserve $Rg_i^t$

Inertia $H_k^t$
Damping $D_k^t$
Power $P_k^t$
Reserve $Re_k^t$

RES     Generator     ESS

Fig. 6. The framework of the proposed FCS-LAPD model.

In this section, the proposed FCS-LAPD model is tested on modified IEEE 24-bus system and a real power system in China. The case studies were carried out on a laptop with an Intel Core i7-10875H CPU and 16 GB RAM. The programs were developed using Matlab R2021b and the QP model was solved by Gurobi.

To evaluate the performance of the proposed method, two look-ahead power dispatch models with 15-minute resolution were carried out during a day: i) with fixed frequency control parameters; ii) the proposed online allocation method.

The nominal frequency $f_0$ is 50 Hz, the maximum allowable frequency deviation $f_{max}^{lim}$, RoCoF $\dot{f}^{lim}$, and steady-state deviation $f_{ss}^{lim}$ are 0.5 Hz, 0.5 Hz/s and 0.25 Hz, respectively. All of the allowable probabilities $\alpha_{Gu}, \alpha_{Gd}, \alpha_{Rw}, \alpha_{L+}, \alpha_{L-}$ are set to 0.05. The size of the initial samples of the CHA method is $5 \times 10^4$. The probability distribution of RES is modeled by GMM, which is fitted based on historical data of wind farms from southern China power grid. Table II and III show the frequency regulation parameters of thermal generators and the parameter value/range of RESs and ESSs.

TABLE II
FREQUENCY REGULATION PARAMETERS OF THERMAL GENERATORS

|  | Inertia $H$ (s) | Droop factor $1/R$ (p.u.) | Fraction $F_H$ (p.u.) |
|---|---|---|---|
| Thermal generator | 4 - 7.5 | 15 - 30 | 0.15 - 0.30 |

TABLE III
PARAMETER VALUE OR RANGE OF RESs AND ESSs

| Method | | Virtual inertia $H$ (s) | Droop coefficient $D$ (p.u.) |
|---|---|---|---|
| RES | Fixed value | 2 | 5 |
|  | Online allocation | 0 - 5 | 0 - 10 |
| ESS | Fixed value | 4 | 10 |
|  | Online allocation | 0 - 5 | 0 - 15 |

### A. Modified IEEE 24-Bus System

The modified IEEE 24-bus system contains 23 thermal generators, 34 transmission lines, and 3 wind farms. The total capacity of wind farms is 700 MW installed with 300 MW/1200 MWh ESS, and the power disturbance $\Delta P^t$ is set to 0.15 of the current load demand.

#### 1) Comparison of Results of Two Power Dispatch Models:

The load demand and the RES forecast power are shown in Fig.7, where the RES reserves of two dispatch models are also demonstrated. The RES reserve almost keeps constant during one day with fixed frequency control parameters, but it varies with the load demand level after online allocation.

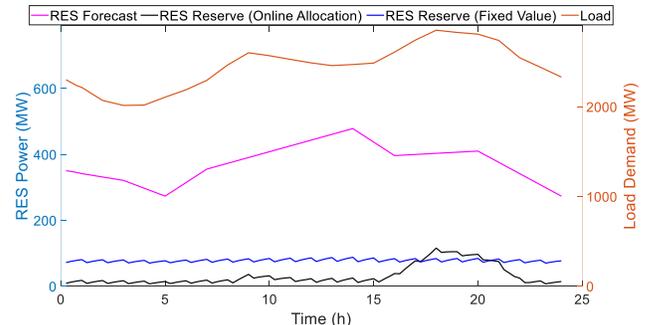

Fig. 7. Load demand and RES power of IEEE 24-bus system.

Table IV shows the results of the two power dispatch models, and the allocation of the inertia and damping (droop coefficient) of RESs and ESSs during one day are illustrated in Fig. 8.

TABLE IV
RESULTS OF TWO DISPATCH MODELS OF IEEE 24-BUS SYSTEM

|  | Fixed value | Online allocation |
|---|---|---|
| Fuel cost ($) | 4100365 | 4037645 |
| RES reserve cost ($) | 949262 | 276526 |
| RES curtailment | 20.78% | 8.53% |
| ESS reserve cost ($) | 23328 | 23488 |

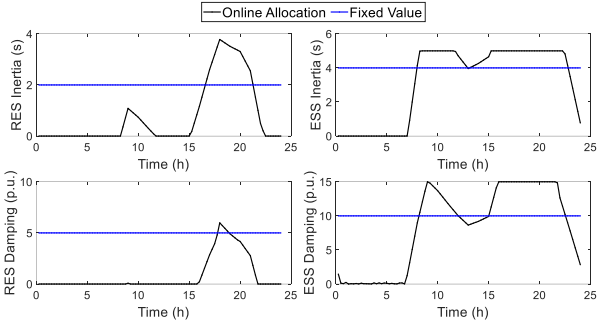

Fig. 8. Allocation of inertia and damping in IEEE 24-bus system.

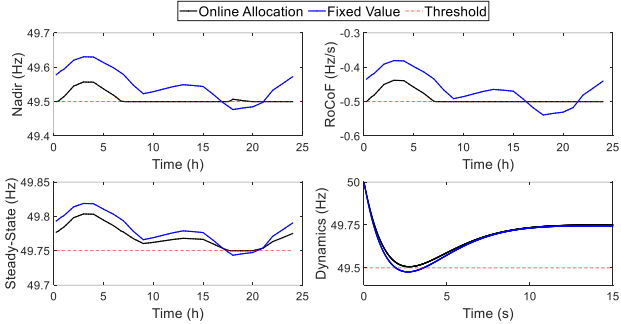

Fig. 9. Frequency dynamic metrics in IEEE 24-bus system.

Compared with fixed frequency control parameters, the online allocation method can obtain a result with lower operating cost and less RES curtailment. This is because the inertia and damping of RESs/ESSs are online allocated according to the realistic frequency support requirement. During periods with low load level (e.g. early morning), the frequency supports from thermal generators are enough, but during high load periods (e.g. evening), more frequency supports are required, so that RESs/ESSs need to participate in frequency regulation. Due to the cost coefficients set as (36), the priority of ESSs to provide frequency support is higher than that of RESs, as shown in Fig. 8.

Fig. 9 shows the frequency dynamic metrics under imaginary disturbance $\Delta P^t$ during one day, and the lower right subgraph illustrates the frequency dynamic response at 18:00. For fixed frequency control parameters method in periods with high load level (e.g. evening), the frequency supports from RESs/ESSs are insufficient, and the system frequency is unsecure. Therefore, the online optimization for frequency control parameters of RESs/ESSs is necessary in power systems with high RES penetration since it can reduce the operation cost under the premise of stabilizing system frequency.

*2) Influence of RES Forecast Precision:*

In this paper, the virtual inertia, droop coefficient and reserve are optimized in look-ahead stage, because the forecast precision of RESs is higher than that in day-ahead stage. The low forecast precision would cause to a too conservative solution due to the chance constraint (25). The FCS-LAPD model is solved twice under different RES forecast precision (with same expectation but different variance), as shown in Fig. 10. The dashed lines (90% range) are 0.05-quantile, 0.95-quantile of total RES power. Due to the forecast uncertainty, more RES power would be curtailed under constraint (25), but the RES curtailment could be reduced by increasing forecast precision. During each hour, the prediction accuracy of RESs gradually decreases, so that the required RES reserve increases. Therefore, the reserve always decreases immediately after the next power dispatch updates, which shows a zigzag shape, as shown in Fig. 7 and Fig. 10.

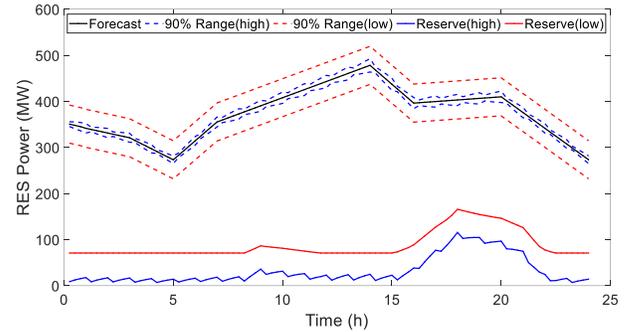

Fig. 10. RES power under different forecast precision of IEEE 24-bus system.

### B. Provincial Power System in China

The real provincial power system of China contains 319 buses, 431 transmission lines, 37 thermal generators, and 22 wind farms. The total capacity of RESs is 3200 MW installed with 1200 MW/4800 MWh ESS, and the power disturbance $\Delta P$ is set to 0.14 of the current load demand.

The load demand and the RES forecast power/reserve are shown in Fig.11. Table V shows the results of the two power dispatch models, and the allocation of the inertia and damping (droop coefficient) of RESs/ESSs are illustrated in Fig. 12. Fig. 13 shows the frequency dynamic metrics under imaginary disturbance $\Delta P^t$ during one day, and the lower right subgraph illustrates the frequency dynamic response at 20:00.

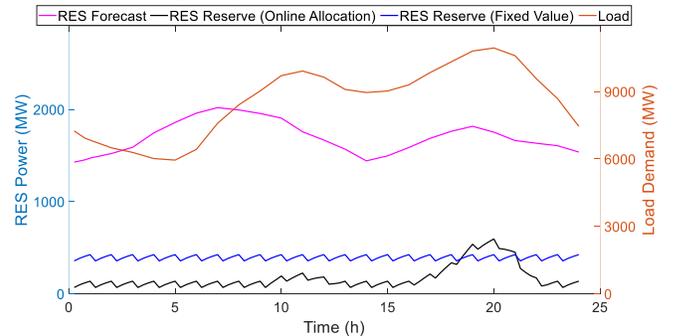

Fig. 11. Load demand and RES power of provincial power system.

TABLE V
RESULTS OF TWO DISPATCH MODELS OF PROVINCIAL POWER SYSTEM

|  | Fixed value | Online allocation |
|---|---|---|
| Fuel cost ($) | 14236379 | 13656769 |
| RES reserve cost ($) | 5287650 | 1543765 |
| RES curtailment | 23.02% | 10.11% |
| ESS reserve cost ($) | 186624 | 158817 |

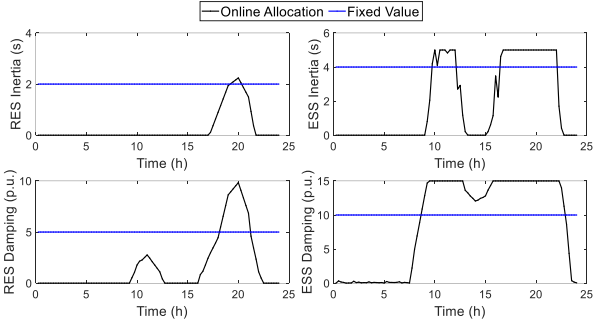

Fig. 12. Allocation of inertia and damping in provincial power system.

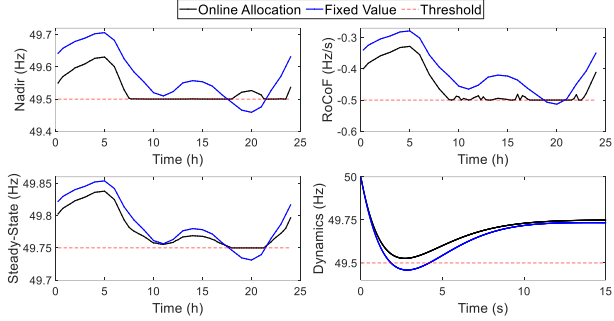

Fig. 13. Frequency dynamic metrics in provincial power system.

It is observed that the proposed online allocation method can achieve lower operating cost and less RES curtailment while stabilizing system frequency compared with fixed frequency control parameter method. The proposed method makes the total system inertia and damping more flexible to cope with power disturbances, which is of importance to the safe and economical operation of high RES penetration power systems.

## V. CONCLUSION

In this paper, a novel frequency constrained stochastic look-ahead power dispatch model is proposed to optimize the frequency control parameters in power scheduling, which can optimally allocate the virtual inertia and droop control of RESs/ESSs, and comprehensively optimize the scheduled power and reserve for all units. The nonlinear frequency nadir constraint is linearized by employing convex hull approximation method. The reserve cost coefficients are adjusted properly to ensure each RES/ESS share the reserve in proportion to its forecast generation/power rating. Case studies are carried out on several test systems. It is shown that the feasible region of frequency constraints can be approximated accurately using CHA method, and the proposed FCS-LAPD model can obtain an optimal result with lower operating cost and less RES curtailment while guaranteeing the frequency dynamic security of power systems.

## APPENDIX I

The maximum frequency deviation $\Delta f_{max}$ is a nonlinear function $F: \mathbb{R}^2 \to \mathbb{R}$.

$$\Delta f_{max} = F(H, D) \tag{30}$$

The function $F$ is twice differentiable, the proof of the convexity of $F$ is equivalent to prove the Hessian matrix $\nabla^2 F$ is positive semidefinite. $\nabla^2 F$ can be expressed as:

$$\nabla^2 F = \begin{bmatrix} \dfrac{\partial^2 F}{\partial H^2} & \dfrac{\partial^2 F}{\partial H \partial D} \\ \dfrac{\partial^2 F}{\partial D \partial H} & \dfrac{\partial^2 F}{\partial D^2} \end{bmatrix} \tag{31}$$

By using the Symbolic Math Toolbox in Matlab [28], we can obtain the symbolic formulation of the Hessian matrix. The formulation is too complex to obtain the parameter range corresponding to the positive semidefinite Hessian matrix. From the opposite point of view, if the Hessian matrix numerically calculated within the actual power system parameter range is a positive semidefinite matrix, it can be proved that $\Delta f_{max}$ is a convex function on $H$ and $D$.

TABLE VI
PARAMETER RANGE OF ACTUAL POWER SYSTEMS

| Parameters | Lower bound | Upper bound |
|---|---|---|
| Total inertia $H$ (s) | 0.1 | 20 |
| Total damping $D$ (p.u.) | 0 | 15 |
| Governor time constant $T$ (s) | 0.1 | 20 |
| Aggregated droop factor $R$ (p.u.) | 1 | 100 |
| Aggregated fraction $F_F$ (p.u.) | 0 | 0.8 |

The parameter range of actual power systems are shown in Table VI. It is worth noting that the parameter $F = F_F \times R$.

In this paper, we perform uniform sampling within the range from Table VI and calculate the Hessian matrices. The correlation between the parameters is not considered in the sampling process ($R$ and $H$ will increase simultaneously as the number of online units increases). If the Hessian matrix without considering correlation is positive semidefinite, the Hessian matrix considering correlation must be positive semidefinite, too. The credibility of the proof increases as the number of samples increases. We did $10^5$ samplings in total, and the obtained Hessian matrices are all positive definite, which indicates that $\Delta f_{max}$ is a strictly convex function on $H$ and $D$.

## APPENDIX II

The allocation of the reserve for RES can be modeled as:

$$\min \sum_{t=1}^{Nt} \left[ \cdots \sum_{j=1}^{Nw} rwc_j^t \left( Rw_j^t \right)^2 \cdots \right]$$

$$\text{s.t.} \sum_{i=1}^{Ng} P_i^t + \sum_{j=1}^{Nw} \left( W_{j,fore}^t - Rw_j^t \right) + \sum_{k=1}^{Ne} P_k^t = \sum_{d=1}^{Nd} D_d^t \tag{32}$$

$$\vdots$$

Assuming all inequality constraints are inactive, the Karush-Kuhn-Tucker (KKT) conditions imply that the optimal solution satisfies:

$$2 rwc_j^t Rw_j^t = \lambda^t \tag{33}$$

where $\lambda^t$ is the multiplier of the equality constraint.

Since the above condition is true for all RES, we get:

$$rwc_j^t Rw_j^t = rwc_{j'}^t Rw_{j'}^t, \forall j, j' \tag{34}$$

We hope each RES can share the reserve in proportion to its forecast generation:

$$\frac{Rw_j^t}{W_{j,fore}^t} = \frac{Rw_{j'}^t}{W_{j',fore}^t}, \forall j, j' \tag{35}$$

Therefore, the cost coefficient $rwc_j^t$ should be inversely proportional to the forecast generation: $rwc_j^t = rwc / W_{j,fore}^t$. If some inequality constraint is active, according to which the reserve will be restricted into a specific value.

The allocation of the reserve for ESS is similar to that of RES. We hope each ESS can share the reserve in proportion to its power rating, so the cost coefficient should be set as $rec_k = rec/P_k^{max}$.

The constant cost coefficient of RESs ($rwc > 0$) and ESSs ($rec > 0$) can be adjusted properly to exhibit the preference of frequency reserve provided by the two different kinds of units. Compared with RES which needs to operate in de-loading mode, the ESS can provide reserve more economically. Therefore, the coefficients $rwc$, $rec$ should satisfy:

$$\frac{rwc}{W_{j,fore}^t} > \frac{rec}{P_k^{max}}, \forall j, k \tag{36}$$

The incremental rates of the objective function $f_{obj}$ for the thermal generators and RESs can be expressed as:

$$\frac{\partial f_{obj}}{\partial P_i^t} = 2a_i P_i^t + b_i > 0 \tag{37a}$$

$$\frac{\partial f_{obj}}{\partial W_{j,sche}^t} = -2rwc_j^t(W_{j,fore}^t - W_{j,sche}^t) < 0 \tag{37b}$$

According to the equal incremental principle, the RES units have a higher priority for generation due to their negative incremental rates, and tend to no curtailment unless necessary frequency regulation reserves are required.